\documentclass{jetpl}
\usepackage{graphicx}
\usepackage{amsmath}
\usepackage{amssymb}
\usepackage[dvipdfm]{hyperref}

\twocolumn

\title{Interaction of two magnetic resonance modes in polar phase of superfluid $^{\bf 3}$He}

\rtitle{Interaction of two magnetic resonance modes in polar phase of superfluid $^3$He}

\sodtitle{Interaction of two magnetic resonance modes in polar phase of superfluid $^3$He}

\author{V.\,V.\,Dmitriev$^+$\footnotemark, A.\,A.\,Soldatov$^{+*}$, A.\,N.\,Yudin$^+$}

\rauthor{V.\,V.\,Dmitriev, A.\,A.\,Soldatov, A.\,N.\,Yudin}

\sodauthor{V.\,V.\,Dmitriev, A.\,A.\,Soldatov, A.\,N.\,Yudin}

\address{$^+$Kapitza Institute for Physical Problems of RAS, 119334 Moscow, Russia}
\address{$^*$Moscow Institute of Physics and Technology, 141700 Dolgoprudny, Russia}

\dates{\today}{*}

\begin{document}

\abstract{
We report results of low frequency nuclear magnetic resonance (NMR) experiments
in the superfluid polar phase of $^3$He which is stabilized by a new type of ``nematic''
aerogel -- nafen. We have found that an interaction between transverse and longitudinal NMR modes may essentially influence the spin dynamics. Theoretical formulas for
NMR resonant frequencies are derived and applied for
interpretation of the experimental results.}

\maketitle

\section{Introduction}
\footnotetext{e-mail: dmitriev@kapitza.ras.ru}
``Nematic'' aerogels (N-aerogels) consist of strands which are oriented along the same direction. Diameters of the strands ($\sim10$\,nm) are less than the superfluid coherence length of $^3$He, so in the case of liquid $^3$He confined in N-aerogel these strands play a role of anisotropic impurities, and theory predicts that superfluid phases which do not exist in bulk $^3$He may become favorable, namely the polar-distorted A phase, the polar-distorted B phase and the pure polar phase \cite{AI,Sauls,Fom,Ik}.
There are two types of N-aerogel: ``Obninsk aerogel'' and nafen \cite{we15}. Polar-distorted A and B phases have been observed in $^3$He confined in Obninsk aerogel \cite{we12,Bph,we14}, while the pure polar phase is realized in $^3$He in nafen \cite{prl}.
As well as in other superfluid phases of $^3$He, in the polar phase two NMR modes may be observed. Frequencies of these modes depend on temperature, on a magnetic field ($\bf H$) and on the orientation of $\bf H$ with respect to the orbital part of the order parameter. Here we present results of continuous wave (CW) NMR experiments in the polar phase of $^3$He in nafen which demonstrate a strong interaction between these two resonant modes at certain conditions. Exact expressions for NMR frequencies in the polar phase in the limit of small excitations are derived and used to explain experimental data.

\section{Theory}
The order parameter of the polar phase is
\begin{equation}
A_{jk} =\Delta_0 e^{i\phi}d_{j}m_{k}, \label{eq1}
\end{equation}
where $\Delta_0$ is the gap parameter, $\phi$ is the phase factor, ${\bf d}$ is the unit spin
vector and ${\bf m}$ is the unit vector in the orbital space which direction
is fixed along the direction of strands of N-aerogel \cite{AI}.
We choose ${\bf H}=H{\bf \hat z}$ and $m_y=0$, so that $m_z=\cos\varphi$ and $m_x=\sin\varphi$, where $\varphi$ is the angle between $\bf H$ and $\bf m$. In the equilibrium state orientation of $\bf d$ is determined by minimization of the sum of dipole ($U_D\propto({\bf dm})^2$) and magnetic ($U_H\propto({\bf dH})^2$) energies, that is ${\bf d}\parallel {\bf \hat y}$.

Spin dynamics of the polar phase is described by Leggett equations \cite{VW}:
\begin{equation}
\begin{array}{lcl}
\dot{\bf M}=\gamma{\bf M}\times{\bf H}-\frac{\Omega^2_P}{\omega_L}\left({\bf d}\times{\bf m}\right)\left({\bf d}{\bf m}\right),\\
\dot{\bf d}={\bf d}\times\left(\gamma{\bf H}-\omega_L{\bf M}\right),\label{Leggett0}
\end{array}
\end{equation}
where $\gamma\approx20378$\,rad/s\,$\cdot$\,Oe is the gyromagnetic ratio of $^3$He, $\omega_L=\gamma H$ is the Larmor frequency, $\bf M$ is the magnetization normalized to its equilibrium value ($\chi H$), $\chi$ is the magnetic susceptibility and $\Omega_P$ is the Leggett frequency of the polar phase, which is zero at the superfluid transition temperature and grows up to $\sim100$\,kHz on cooling. Introducing $\widetilde{M}_z=M_z-1$ Eqs.(\ref{Leggett0}) can be rewritten as follows:
\begin{equation}
\begin{array}{lcl}
\dot{M}_x=\omega_L M_y-\frac{\Omega^2_P}{\omega_L}\left({\bf dm}\right)d_y m_z,\\
\dot{M}_y=-\omega_L M_x-\frac{\Omega^2_P}{\omega_L}\left({\bf dm}\right)\left(d_z m_x-d_x m_z\right),\\
\dot{\widetilde{M}}_z=\frac{\Omega^2_P}{\omega_L}\left({\bf dm}\right)d_y m_x,\\
\dot{d}_x=\omega_L\left(d_z M_y-d_y\widetilde{M}_z\right),\\
\dot{d}_y=\omega_L\left(d_x \widetilde{M}_z-d_z M_x\right),\\
\dot{d}_z=\omega_L\left(d_y M_x-d_x M_y\right).\label{Leggett1}
\end{array}
\end{equation}
We consider small deviations of ${\bf M}$ and ${\bf d}$ from
the equilibrium state, that is: $\widetilde{M}_z\ll 1$, $M_{x,y}\ll 1$, $d_{x,z}\ll 1$, and $d_y\approx1$.
Then from Eqs.(\ref{Leggett1}) it follows:
\begin{equation}
\begin{array}{lcl}
\ddot{M}_x=-\left(\omega^2_L+\Omega^2_P \cos^2\varphi\right)M_x+\Omega^2_P \widetilde{M}_z \sin\varphi \cos\varphi,\\
\ddot{\widetilde{M}}_z=\Omega^2_P M_x\sin\varphi \cos\varphi -\Omega^2_P \widetilde{M}_z \sin^2\varphi,
\label{ddotM}
\end{array}
\end{equation}
that results in the equation for NMR frequencies $\omega$:
\begin{equation}
\omega^4-\left(\omega^2_L+\Omega^2_P\right)\omega^2+\Omega^2_P\omega^2_L\sin^2\varphi=0.
\label{freq}
\end{equation}
Eq.(\ref{freq}) has two solutions:
\begin{equation}
\begin{split}
\omega^2_\pm=\frac{1}{2}\left(\omega^2_L\right.&+\Omega^2_P\pm\\
&\pm\left.\sqrt{\left(\omega^2_L+\Omega^2_P\right)^2-4\omega^2_L\Omega^2_P\sin^2\varphi}\right).
\end{split}
\label{twomodes}
\end{equation}
We note that there is a frequency gap between $\omega_+$ and $\omega_-$ modes: there are no solutions of Eq.(\ref{freq}) for $\Omega_P>\omega>\Omega_P\sin\varphi$. In general, both modes represent coupled oscillations of transverse and longitudinal components of $\bf M$, and the ratio of amplitudes of transverse and longitudinal oscillations for a given $\omega$ is
\begin{equation}
R=\frac{|\omega^2-\Omega^2_P\sin^2\varphi|}{\Omega^2_P\sin\varphi\cos\varphi}.
\label{ratio}
\end{equation}

\begin{figure}[t]
\center
\includegraphics[width=0.95\linewidth]{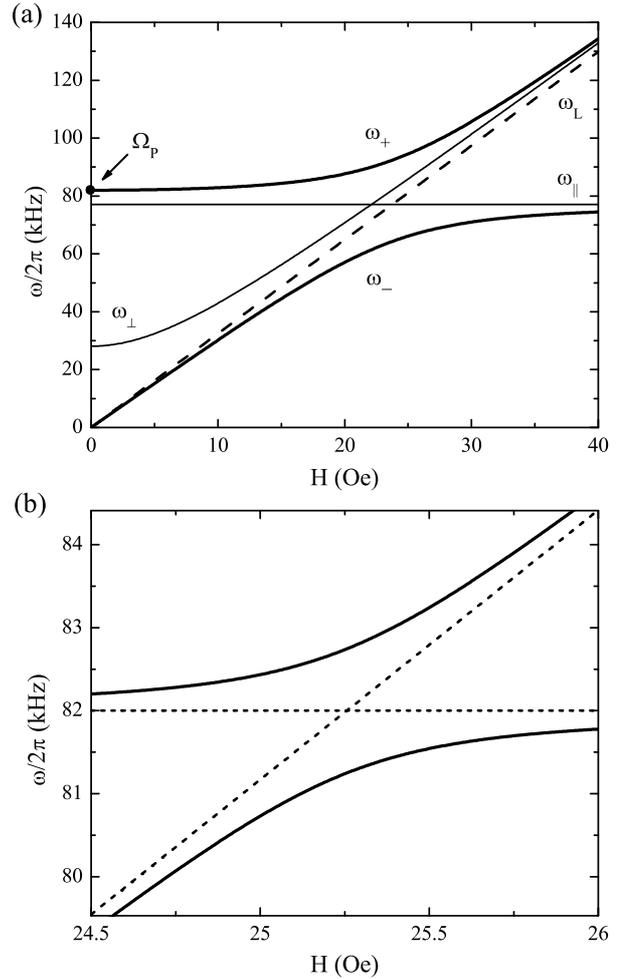}
\caption{Fig. \thefigure: CW NMR frequencies in the polar phase versus H at $\Omega_P/2\pi=82$ kHz.
(a) From Eq.(\ref{twomodes}) (thick solid lines) and from Eqs.(\ref{prev}) (thin solid lines). Dashed line corresponds to the Larmor frequency. $\varphi=68^\circ$.
(b) From Eq.(\ref{twomodes}) for $\varphi=89^\circ$ (solid lines) and for $\varphi=90^\circ$ (short-dashed lines).}
\label{figtwomodes}
\end{figure}

Expressions for NMR frequencies in the polar phase derived in previous papers \cite{AI,we12,prl,zav} were obtained in the assumption that equations for $M_z(t)$ and $M_{x,y}(t)$ are decoupled. In this case we get two noninteracting NMR modes: longitudinal and transverse. Frequencies of these modes for small amplitudes are
\begin{equation}
\begin{array}{lcl}
\omega_{\parallel}=\Omega_P\sin\varphi,\\
\omega_{\perp}=\sqrt{\omega_L^2+\Omega_P^2\cos^2\varphi}.\label{prev}
\end{array}
\end{equation}
Eqs.(\ref{prev}) are valid only for $\varphi=0$ and $\varphi=90^\circ$, or if $\omega_L \gg \Omega_P$. In other cases, as it follows from Eqs.(\ref{ddotM}), the coupling of longitudinal and transverse modes becomes essential. This is illustrated by Fig.\ref{figtwomodes}a, where we present calculated field dependencies of $\omega_+$, $\omega_-$, $\omega_{\parallel}$, and $\omega_{\perp}$ for realistic experimental conditions. It is seen that the coupling between resonant modes results in their ``repulsion'' and transformation into two nonintersecting branches with frequencies $\omega_+$ and $\omega_-$.
We note that at $\varphi=90^\circ$ we get two intersecting modes, but even at small deflections of $\varphi$ from $90^\circ$ a qualitative change of the NMR spectrum occurs (Fig.\ref{figtwomodes}b).

\section{Details of experiment}
In the present work we use the same experimental chamber as in experiments described in \cite{diff}. The chamber is made of Stycast-1266 epoxy resin and has two cells with nafen samples which were produced by ANF Technology Ltd (Tallinn, Estonia). In the experiments described below we use the cell with nafen with overall density of 243\,mg/cm$^3$ (nafen-243). The sample has a cuboid shape with sizes of 4\,mm and is placed freely in the cell. It consists of Al$_2$O$_3$ strands with diameters of $\sim9$\,nm and has a porosity of 93.9\%. More information about the sample can be found in \cite{we15}.

The necessary temperatures were obtained
by a nuclear demagnetization cryostat and measured by a quartz tuning fork calibrated by
measurements of the Leggett frequency in bulk $^3$He-B. To avoid a
paramagnetic signal from surface solid $^3$He, the sample was preplated by $\sim2.5$ atomic
monolayers of $^4$He.

Experiments were performed using transverse CW NMR in magnetic fields of $25\div111$\,Oe (corresponding
NMR frequencies were $82\div360$ kHz) and at a pressure of 29.3\,bar.
The superfluid phase diagram of $^3$He in nafen-243 is presented in \cite{prl}. At 29.3\,bar the superfluid transition temperature of $^3$He in nafen-243 is suppressed by $\sim2$\% with respect to the superfluid transition temperature ($T_c$) in bulk $^3$He, and down to the lowest reached temperature ($\sim 0.47\,T_c$) the only observed superfluid phase is the polar phase.

Two solenoids were used in order to apply the external magnetic field in directions parallel (${\bf H}_\parallel$,
longitudinal field) and perpendicular (${\bf H}_\perp$, transverse field) to nafen strands.
So, the resultant field ${\bf H}={\bf H}_\parallel+{\bf H}_\perp$ could be rotated by an arbitrary
angle $\varphi$ with respect to the anisotropy axis of nafen. CW NMR measurements in the normal phase of $^3$He show that an angle between axes of longitudinal and transverse solenoids is 90$^{\circ}\pm 0.2^\circ$. However, we estimate an accuracy of setting $\varphi$ as $\pm 1^\circ$ due to a possible misalignment between the axis of the longitudinal solenoid and the anisotropy axis of the nafen sample. For $H_\perp\sim25$\,Oe it limits an accuracy in determining of $H_\parallel$ to 0.4\,Oe.

\section{Results}
Experiments were carried out using transverse CW NMR at a fixed frequency $\omega$ for $\varphi \approx 68^\circ$ and for $\varphi$ close to $90^\circ$. We applied fixed field ${\bf H}_\parallel$ and swept ${\bf H}_\perp$ to record the NMR line. In terms of $H_\parallel$ and $H_\perp$ Eq.(\ref{freq}) can be rewritten as follows:
\begin{equation}
\left(\gamma H_\perp\right)^2=\omega^2\left[1-\left(\gamma H_\parallel\right)^2/\left(\omega^2-\Omega^2_P\right)\right].
\label{freqH}
\end{equation}
In this case $H_\perp/H_\parallel=\tan\varphi\ne const$, but
the change of $H_\perp$ during the sweep through the NMR line is small, and the corresponding change of $\varphi$ is less than $0.1^\circ$.

\begin{figure}[t]
\center
\includegraphics[width=0.95\linewidth]{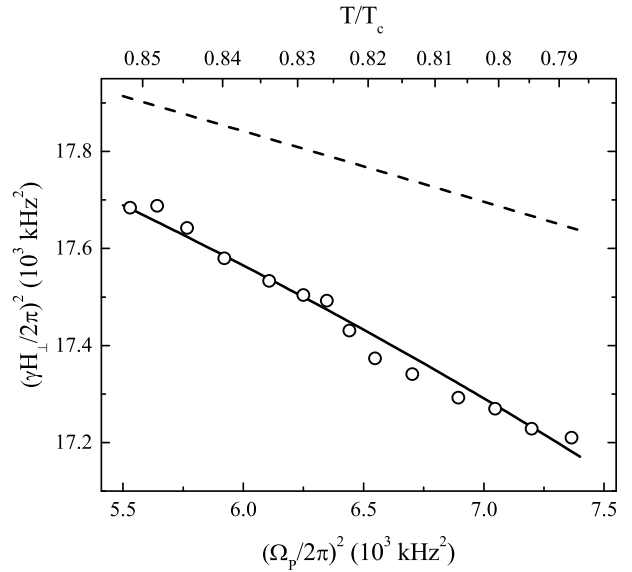}
\caption{Fig. \thefigure: CW NMR transverse magnetic field in the polar phase as a function of temperature (upper scale) and $\Omega_P$ (lower scale). Open circles are
experimental data at $\omega/2\pi=146.8$\,kHz, $H_\parallel\approx16.4$ Oe. $H_\perp$ is in the range of $40.4\div41.0$\,Oe, so that
$\varphi$ is in the range of $67.9^\circ \div 68.2^\circ$. Solid curve is best fit by Eq.(\ref{freqH}) with a single fit
parameter of $H_\parallel$ which was found to be equal to 16.53\,Oe. Dashed curve corresponds to Eq.(\ref{prev}) rewritten in terms of $H_\parallel$ and $H_\perp$ with $H_\parallel=16.53$\,Oe. Temperature dependence of $\Omega_P$ was measured independently by CW NMR at $\varphi=0$.}
\label{phi70}
\end{figure}

\begin{figure}[t]
\center
\includegraphics[width=0.95\linewidth]{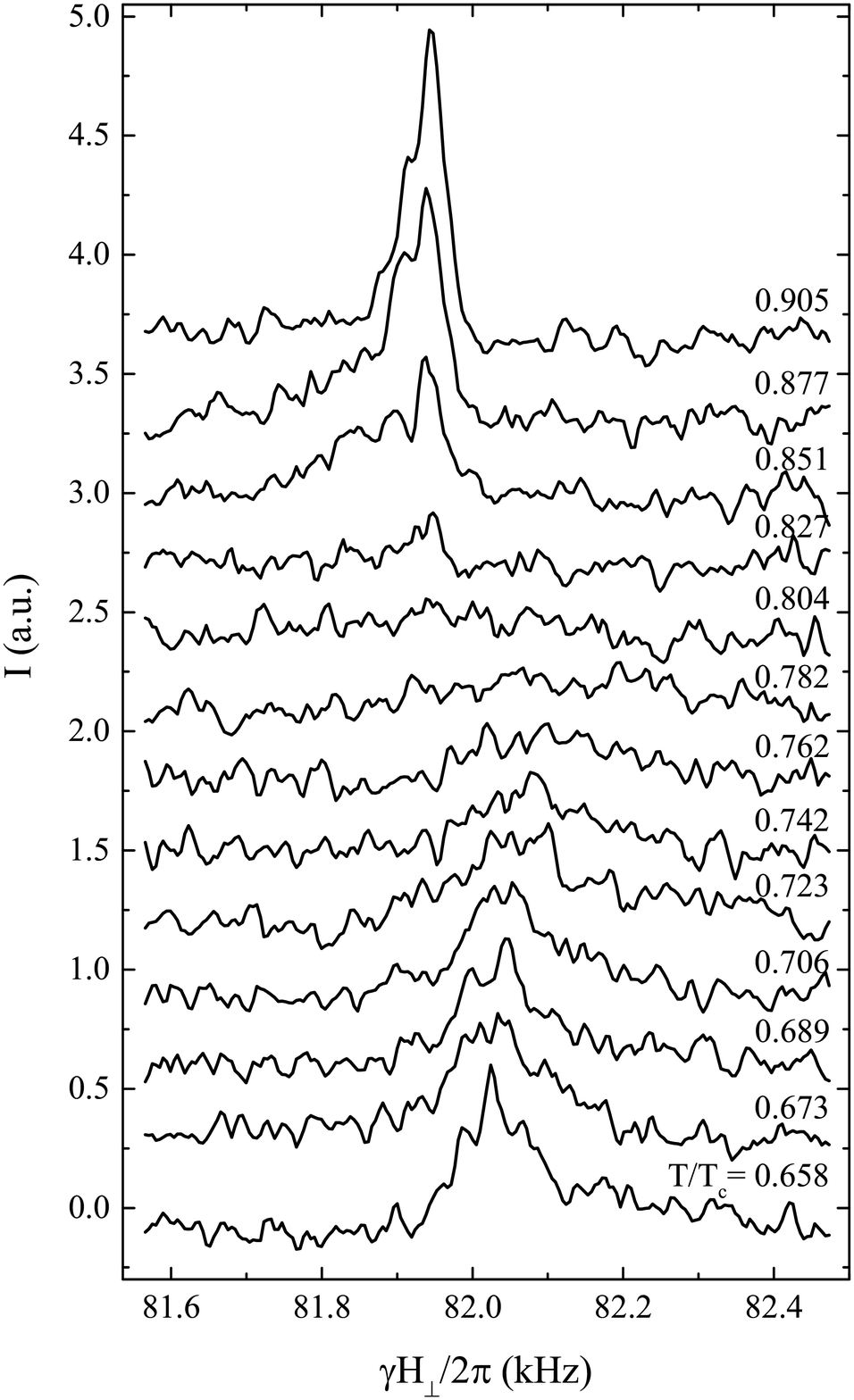}
\caption{Fig. \thefigure: CW NMR absorption lines recorded during warming in the polar phase. The lines at different temperatures are shifted in vertical direction for a better view. $H_\parallel$ was set to 0.29\,Oe, but further analysis (see Fig.\ref{phi90}) pointed out that it corresponds to 0.54\,Oe ($\varphi\approx88.8^\circ$) that may be due to misalignment between the axis of the longitudinal solenoid and the direction of nafen strands.
$\omega/2\pi=82$\,kHz, $H\approx25$\,Oe.}
\label{NMRphi90}
\end{figure}
\begin{figure}[t]
\center
\includegraphics[width=0.95\linewidth]{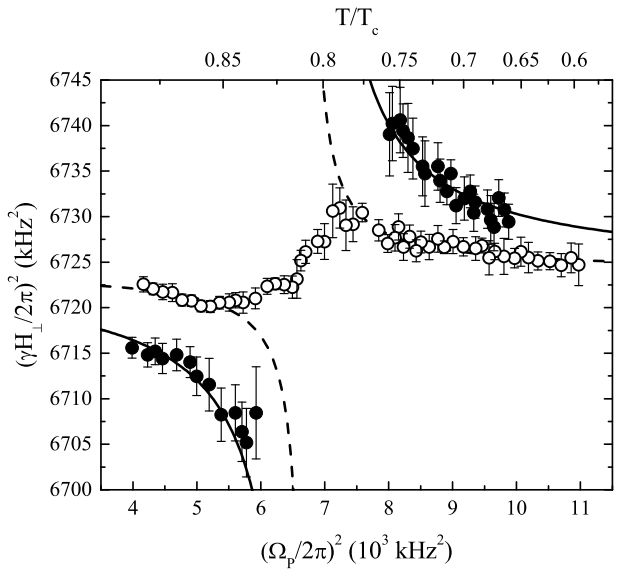}
\caption{Fig. \thefigure: The dependence of $H_\perp$ on temperature (upper scale) and $\Omega_P$ (lower scale). $H_\parallel$ was set to 0.29\,Oe (filled circles) and to 0.17\,Oe (open circles). Lines are best fits of the data by Eq.(\ref{freqH}) with $H_\parallel=0.54$\,Oe (solid line) and $H_\parallel=0.27$\,Oe (dashed line). $\omega/2\pi=82$\,kHz, $H\approx25$\,Oe.}
\label{phi90}
\end{figure}

The measured dependence of the resonant field $H_\perp$ on temperature (and $\Omega_P$) at $\varphi\approx68^\circ$ is shown in Fig.\ref{phi70}. It is seen that the experimental data are in a good agreement with Eq.(\ref{freqH}) (solid curve) which fits the data much better than Eq.(\ref{prev}) (dashed curve). We note that small variations of $H_\parallel$ in the equations result only in vertical shift of the theoretical curves (by $\sim 100$\,kHz$^2$ for $\delta H_\parallel=0.2$\,Oe), but their slopes practically remain the same. Therefore, we think that a small difference between set and fitted values of $H_\parallel$ is due to the inaccuracy in setting of $H_\parallel$ due to the above mentioned misalignment.

In experiments presented in Fig.\ref{phi70} only the mode with frequency $\omega_+$ was excited because $\omega$ was always greater than $\Omega_P$. The transition between $\omega_+$ and $\omega_-$ modes can be observed if at a fixed $\omega$ the temperature (and consequently $\Omega_P$) is changed, so that we cross the frequency gap between modes. Maximal value of $\Omega_P$ which we could obtain at $T\approx 0.47$\,$T_c$ was 107\,kHz. Therefore, in order to observe the transition between modes we used the NMR frequency of 82\,kHz, although our NMR setup was not optimal for such a low frequency, and the signal-to-noise ratio was rather poor. It is also worth noting that an absolute value of $\partial\omega/\partial H_\bot$ decreases if $\omega$ is approaching $\Omega_P$ (or  $\Omega_P\sin\varphi$) resulting in a broadening of the NMR line and in an additional decrease of the signal-to-noise ratio near the transition region. For these reasons we were able to see the transition between the modes only if $\varphi$ is close to 90$^\circ$ where the frequency gap is small enough. Transverse CW NMR absorption lines recorded at $\varphi\approx89^\circ$ during slow warming in the polar phase are shown in Fig.\ref{NMRphi90}. The transition between $\omega_+$ and $\omega_-$ modes occurs in a narrow temperature region near $\sim0.8$\,$T_c$ where the NMR signal practically disappears. At higher temperatures we excite $\omega_+$-mode while at lower temperatures $\omega_-$-mode is observed.

The dependence of the resonant field $H_\perp$ on $\Omega_P$ (calculated from the first moment of the NMR line) is shown in Fig.\ref{phi90} by filled circles. It agrees well with the dependence following from Eq.(\ref{freqH}) if we assume that the real value of $H_\parallel$ equals 0.54\,Oe (solid curve).

We also have done a similar experiment where we set $H_\parallel$ to be equal to 0.17\,Oe (open circles in Fig.\ref{phi90}). In this case the NMR signal was absent in essentially smaller range of temperatures (between $0.79\,T_c$ and $0.81\,T_c$) and at temperatures $0.81\,T_c<T<0.83\,T_c$ the experimental data cannot be fitted by Eq.(\ref{freqH}). We assume that this is explained by variations of $\varphi$ inside the sample about the mean value. If we do not take into account the data in this temperature range then best fit by Eq.(\ref{freqH}) corresponds to $H_\parallel=0.27$\,Oe (i.e. $\varphi \approx 89.4^\circ$), and the variations of $\varphi$ may be estimated as $\sim 0.6^\circ$.

To sum up, our results of low frequency transverse CW NMR experiments in the polar phase of $^3$He in nafen are in a good agreement with the developed theoretical model. Experiments at $\varphi\approx68^\circ$
(Fig.\ref{phi70}) confirm the validity of Eqs.(\ref{twomodes},\ref{freqH}),
while experiments at angles $\varphi$ close to $90^\circ$ (Fig.\ref{NMRphi90} and
Fig.\ref{phi90}) prove the existence of two non-intersecting branches of the NMR spectrum.

\section{Conclusions}

It was observed for the first time that in the polar phase of $^3$He at even small deviations of $\varphi$ from 90$^\circ$ the longitudinal and transverse NMR modes become coupled that results in their repulsion and transformation into two non-intersecting branches of the NMR spectrum. The coupling between these modes is possible due to N-aerogel which fixes $\bf m$ along the aerogel strands. In contrast, in equilibrium homogeneous state in bulk superfluid $^3$He the coupling of longitudinal and transverse NMR modes cannot be observed because the magnetic field orients order parameters of A and B phases so that the modes do not interact.
However, the coupling may appear in the presence of inhomogeneities of the order parameter, caused by boundaries or textural defects (solitons or vortices). So, a decay of the transverse mode into two longitudinal modes with nonzero wave vectors, as well as the decay of the transverse mode due to emission of short wave acoustic magnons in the presence of quantized vortices, has been observed in recent experiments in the B phase \cite{ZavyalovAutti2016}.

The phenomenon we discuss is also known in other systems with interacting resonant modes.
In particular, our results are quite similar to results of experiments in MnCO$_3$ \cite{mnco3} where field dependencies of two AFM resonant modes were measured for different orientations of a magnetic field.

\section{Acknowledgements}
We are grateful to to G.E.\,Volovik for useful comments and I.M. Grodnensky for providing samples of nafen. This work was
supported in part by RFBR grant 16-02-00349 and the Basic Research Program of the Presidium
of Russian Academy of Sciences.

\end{document}